\documentstyle[]{article}
\begin{document}
\title{Fuzzy, Non Commutative SpaceTime: A New Paradigm for A New Century}
\author{B.G. Sidharth\\
B.M. Birla Science Centre, Adarshnagar, Hyderabad - 500 063, India}
\date{}
\maketitle
\begin{abstract}
Much of twentieth century physics, whether it be Classical or Quantum, has
been based on the concept of spacetime as a differentiable manifold. While
this work has culminated in the standard model, it is now generally accepted
that in the light of recent experimental results, we have to go beyond the
standard model. On the other hand Quantum SuperString Theory and a recent
model of Quantized Spacetime in which, for example, an electron can be
meaningfully described by the Kerr-Newman metric, have shown promise. They
lead to mathematically identical spacetime-energy-momenta commutation
relations, and infact an identical non commutative geometry which is a
departure from the usual concept of spacetime. This could well be a new
paradigm for the new century.
\end{abstract}
\section{Introduction}
At the beginning of the twentieth century several Physicists including Poincare
and Abraham amongst others were tinkering unsuccessfully with the problem of
the extended electron\cite{r1,r2}. The problem was that an extended electron
appeared to contradict Special Relativity, while on the other hand, the limit
of a point particle lead to inexplicable infinities. These infinities dogged
physics for many decades. Infact the Heisenberg Uncertainity Principle straightaway
leads to infinities in the limit of spacetime points. It was only through
the artifice of renormalization that 't Hooft could finally circumvent this
vexing problem, in the 1970s (Cf. paper by 't Hooft in this volume).\\
Nevertheless it has been realized that the concept of spacetime points is only
approximate\cite{r3,r4,r5,r6,r7}. We are beginning to realize that it may be
more meaningful to speak in terms of spacetime foam, strings, branes, non
commutative geometry, fuzzy spacetime and so on\cite{r8}. This is what we
will now discuss.
\section{Two Approaches}
We now consider the well known theory of Quantum SuperStrings and also an
approach in which an electron is considered to be a Kerr-Newman Black Hole,
with the additional input of fuzzy spacetime.\\
As is well known, String Theory originated from phonenomenological considerations
in the late sixties through the pioneering work of Veneziano, Nambu and
others to explain features like the s-t channel dual resonance scattering and
Regge trajectories\cite{r9}. Originally strings were conceived as one
dimensional objects with an extension of the order of the Compton wavelength,
which would fudge the point vertices of the s-t channel scattering graphs,
so that both would effectively correspond to one another (Cf.ref.\cite{r9}).\\
The string itself is governed by the equation\cite{r10}
\begin{equation}
\rho \ddot{y} - Ty'' = 0\label{e1}
\end{equation}
where the frequency $\omega$ is given by
\begin{equation}
\omega = \frac{\pi}{2} \sqrt{\frac{T}{\rho}}\label{e2}
\end{equation}
\begin{equation}
T = \frac{mc^2}{l}; \rho = \frac{m}{l}\label{e3}
\end{equation}
\begin{equation}
\sqrt{T/\rho} = c\label{e4}
\end{equation}
$T$ being the tension of the string, which has to be introduced in the theory,
$l$ its length and $\rho$ the line density.
The identification (\ref{e3}) gives (\ref{e4}) where $c$ is the velocity of light,
and (\ref{e1}) then goes over to the usual d'Alembertian or massless Klein-Gordon equation. It is worth
noting that as $l \to 0$ the potential energy which is $\sim \int^l_0 T \left(\frac{\partial y}
{\partial x}\right)^2 dx$ rapidly oscillates.\\
Quantization of the states leads to
\begin{equation}
\langle \Delta x^2 \rangle \sim l^2\label{e5}
\end{equation}
The string effectively shows up as an infinite collection of Harmonic Oscillators
\cite{r10}.
It follows from the above that the length $l$ of the string turns out to be the
Compton wavelength, a circumstance which has been described as one of the miracles
of String Theory by Veneziano\cite{r11}.\\
The above strings are really Bosonic strings. Raimond\cite{r12}, Scherk\cite{r13}
and others laid the foundation for the theory of Fermionic strings. Essentially
the relativistic Quantized String is given a rotation, when we get back the
equation for Regge trajectories,
\begin{equation}
J \leq (2\pi T)^{-1}M^2 + a_0\hbar \quad \mbox{with}\quad a_0 = +1(+2)
\mbox{for the open (closed) string}\label{e6}
\end{equation}
Attention must be drawn to the additional term $a_0$ which now appears in (\ref{e6}). It arises
from a zero point energy effect. When $a_0 = 1$ we have gauge Bosons while
$a_0 = 2$ describes the gravitons. In the full theory of Quantum Super Strings,
we are essentially dealing with extended objects rotating with the velocity of
light, rather like spinning black holes. The spatial extention is at the
Planck scale while features like extra space time dimensions which are curled
up in the Kaluza Klein sense and, as we will see, non commutative geometry
appear\cite{r14,r15}.\\
The above considerations raise the question, can a charged elementary particle be
pictured as a Kerr Newman Black Hole, though in a Quantum Mechanical context
rather than the General Relativistic case? Indeed it is well known that the
Kerr Newman Black Hole itself mimics the electron remarkably well including
the purely Quantum Mechanical anomalous $g=2$ factor\cite{r16}. The problem is
that there would be a naked singularity, that is the radius would become complex,
\begin{equation}
r_+ = \frac{GM}{c^2} + \imath b, b \equiv \left(\frac{G^2Q^2}{c^8} + a^2 -
\frac{G^2M^2}{c^4}\right)^{1/2}\label{e7}
\end{equation}
where $a$ is the angular momentum per unit mass.\\
This problem has been studied in detail by the author in recent years\cite{r1,r17,r18}.
Indeed it is quite remarkable that the position coordinate of an electron in
the Dirac theory is non Hermitian and mimics equation (\ref{e7}), being
given by
\begin{equation}
x = (c^2 p_1 H^{-1}t+a_1) + \frac{\imath}{2} c\hbar (\alpha_1 - cp_1H^{-1})H^{-1},\label{e8}
\end{equation}
where the imaginary parts of (\ref{e7}) and (\ref{e8}) are both of the order
of the Compton wavelength.\\
The key to understanding the unacceptable imaginary part was given by Dirac
himself\cite{r19}, in terms of Zitterbewegung. The point is that according to
the Heisenberg Uncertainity Principle, space time points themselves are not
meaningful- only space time intervals have meaning, and we are really speaking of
averages over such intervals, which are atleast of the order of the Compton
scale. Once this is kept in mind, the imaginary term disappears on averaging
over the Compton scale.\\
Indeed, from a classical point of view also, in the extreme relativistic
case, as is well known there is an extension of the order of the Compton
wavelength, within which we encounter meaningless negative energies\cite{r20}.
With this proviso, it has been shown that we could think of an electron as a
spinning Kerr Newman Black Hole. This has received independent support from
the work of Nottale\cite{r21}.
\section{Non Commutative Geometry}
We are thus lead to the picture where there is a cut off in space time intervals
as indicated in the introduction.\\
In the above two scenarios, the cut off is at the Compton scale $(l,\tau)$.
Such discrete space time models compatible with Special Relativity have been
studied for a long time by Snyder and several other scholars\cite{r22,r23,r24}.
In this case it is well known that we have the following non commutative
geometry
$$[x, y] = (\imath a^2 / \hbar)L_{z,}  [t, x] = (\imath a^2 / \hbar c)M_{x,}$$
\begin{equation}
[y, z] = (\imath a^2 / \hbar) L_{x,} [t, y] = (\imath a^2 / \hbar c)M_{y,}\label{e9}
\end{equation}
$$[z, x] = (\imath a^2 / \hbar) L_{y,} [t, z] = (\imath a^2 / \hbar c)M_{z,}$$
where $a$ is the minimum natural unit and $L_x, M_x$ etc. have their usual significance.\\
Moreover in this case there is also a correction to the usual Quantum Mechanical
commutation relations, which are now given by
$$[x, p_x] = \imath \hbar [1+(a/\hbar)^2 p^2_x];$$
$$[t, p_t] = \imath \hbar [1-(a/ \hbar c)^2 p^2_t];$$
\begin{equation}
[x, p_y] = [y, p_x] = \imath \hbar (a/ \hbar)^2 p_xp_y ;\label{e10}
\end{equation}
$$[x, p_t] = c^2[p_{x,} t] = \imath \hbar (a/ \hbar)^2 p_xp_t ;\mbox{etc}.$$
where $p^\mu$ denotes the four momentum.\\
In the Kerr Newman model for the electron alluded to above (or generally
for a spinning sphere of spin $\sim \hbar$ and of radius $l$), $L_x$ etc. reduce
to the spin $\frac{\hbar}{2}$ of a Fermion and the commutation relations
(\ref{e9}) and (\ref{e10}) reduce to
\begin{equation}
[x,y] \approx 0(l^2),[x,p_x] = \imath \hbar [1 + \beta l^2], [t,E] = \imath \hbar [1+\tau^2]\label{e11}
\end{equation}
where $\beta = (p_x/\hbar)^2$ and similar equations.\\
Interestingly the non commutative geometry given in (\ref{e11}) can be shown
to lead to the representation of Dirac matrices and the Dirac equation\cite{r25}.
From here we can get the Klein Gordon equation, as is well known\cite{r26,r27},
or alternatively we deduce the massless string equation (\ref{e1}), using
(\ref{e4}).\\
This is also the case with superstrings where Dirac spinors are
introduced, as indicated in Section 2. Infact in QSS also we have equations
mathematically identical to the relations (\ref{e11}) containing
momenta. This, which implies (\ref{e9}), can
now be seen to be the origin of non-commutativity.\\
The non commutative geometry and fuzzyness is contained in (\ref{e11}).
Infact fuzzy spaces have been investigated in detail by Madore and others\cite{r28,r29},
and we are lead back to the equation (\ref{e11}). The fuzzyness which is
closely tied up with the non commutative feature is symptomatic of the breakdown
of the concept of the spacetime points and point particles at small scales
or high energies. As has been noted by Snyder, Witten, and several other
scholars, the divergences encountered in Quantum Field Theory are symptomatic
of precisely such an extrapolation to spacetime points and which necessitates
devices like renormalization. As Witten points out\cite{r30}, "in developing
relativity, Einstein assumed that the space time coordinates were Bosonic;
Fermions had not yet been discovered!... The structure of space time is
enriched by Fermionic as well as Bosonic coordinates."\\
Interestingly, starting from equation (\ref{e11}), we can deduce that $l$ is
the Compton wavelength without however assuming it to be so. Let us write the
first equation of (\ref{e11}) as
\begin{equation}
[x,y] = \imath H\label{e12}
\end{equation}
The relation (\ref{e12}) shows that $y$ plays a role similar to the $x$ component
of the momentum, and infact mathematically we have
\begin{equation}
y = \frac{\imath H}{\hbar} p_x \equiv \tilde h p_x\label{e13}
\end{equation}
At the extreme energies and speeds, we would have
\begin{equation}
y = \tilde h p = \tilde h mc, m \dot y = p_y, x = \tilde h p_y\label{e14}
\end{equation}
From (\ref{e13}) it follows that
$$y = H \frac{d}{dx}$$
whence
\begin{equation}
Ty'' = \frac{T}{H} y y' = \frac{T}{H^2}y \cdot y^2\label{e15}
\end{equation}
Further from (\ref{e14}) it follows that
\begin{equation}
\rho \ddot y = \frac{\rho}{m} \frac{d}{dt} \left(\frac{x}{\tilde h}\right) =
\frac{\rho}{m^2\tilde h} \cdot \frac{y}{\tilde h} = \frac{\rho}
{m^2 \tilde h^2} y\label{e16}
\end{equation}
Fusing (\ref{e15}) and (\ref{e16}) in to one we get
$$\frac{H^2}{\tilde h^2} \frac{1}{m^2c^2} \equiv l^2 = \left(\frac{h}{mc}\right)^2 = y^2$$
where $l$ is now the Compton wavelength. This is the explanation
for the so called miraculous emergence of the Compton wavelength in string theory, as
noted by Veneziano (Cf.ref.\cite{r11}).\\
Finally it may be pointed out that the tension $T$ of String Theory, appears as
the energy of the Quantum Mechanical Kerr Newman Black Hole alluded to in
Section 2 via the relation (\ref{e3}).\\
We next have to see how, from the Compton scale above, we arrive at the Planck
scale of QSS. For this, we note that from (\ref{e11}), using
$$\Delta p \cdot \Delta x \approx h,$$
we get,
$$\Delta p \cdot \Delta x = \hbar [1+\frac{l^2}{(\Delta x)^2}]$$
Whence
\begin{equation}
\Delta p (\Delta x)^3 = \hbar [(\Delta x)^2 + l^2]\label{e17}
\end{equation}
Witten describes it as an extra correction to the Heisenberg Uncertainity Principle.
As long as we are at usual energies, we have the usual Uncertainity Principle, and
the usual bosonic or commutative spacetime. At high energies however we encounter
the extra term in (\ref{e17}) viz., $\hbar' = \hbar l^2$. With this, the
Compton scale goes over from $l$ to $l^3$, the Planck scale (Cf. also \cite{r31}).
Equally interesting is the fact that as can be seen from (\ref{e17}), the
single $x$ dimension gets trebled. At these Planck scales, therefore, a total
of six extra dimensions appear, which are curled up in the Kaluza Klein sense
at the Planck scale. This provides an explanation for the puzzling six extra dimensions
of QSS.
\section{Further Issues}
{\large {\bf (i) Vortices}}\\ \\
As described in detail in \cite{r17} the Quantum Mechanical Kerr-Newman Black
Hole could also be considered to be a vortex. If we take two parallel spinning
vortices separated by a distance $d$ then the angular velocity is given by
$$\omega = \frac{\nu}{\pi d^2},$$
where $\nu = h/m$.\\
Whence the spin of the system turns out to be $h$, that is in usual units the
spin is one, and the above gives the states $\pm 1$.\\
There is also the case where the two above vortices are anti parallel. In this
case there is no spin, but rather there is the linear velocity given by
$$v = \nu / 2\pi d$$
This corresponds to the state $0$ in the spin $1$ case.\\
Together, the two above cases give the three $-1,0,+1$ states of spin $1$ as in
the Quantum Mechanical Theory.\\
In the case of the Quantum Mechanical Kerr-Newman Black Hole hydrodynamical vortex
pictured above, it is interesting that for the bound state, there is really no
interaction in the particle physics sense. The interaction comes in because in
the above description we really identify a background Zero Point Field with the
hydrodynamical flow (Cf.ref.\cite{r18} and also\cite{r32}). Interestingly in
a simulation involving vortices, such an "attraction" was noticed\cite{r33}.\\ \\
{\large {\bf (ii) Monopoles}}\\ \\
It is interesting that the above considerations lead to a characterization of
the elusive monopole. Infact a non commutative geometry can be associated with
a powerful magnetic field\cite{r34}, and specialising to the equations (\ref{e11})
we can show that this field $B$ satisfies,
$$Bl^2 \sim \frac{nhc}{2e}$$
which is the celebrated equation of the monopole.\\ \\
{\large {\bf (iii) Duality}}\\ \\
A related concept, which one encounters also in String Theory is Duality. Infact
the relation (\ref{e11}) leads to (Cf. also equation (\ref{e17}),
\begin{equation}
\Delta x \sim \frac{\hbar}{\Delta p} + \alpha' \frac{\Delta p}{\hbar}\label{e18}
\end{equation}
where $\alpha' = l^2$, which in Quantum SuperStrings Theory $\sim 10^{-66}$. Witten
has wondered about the basis of (\ref{e18}), but as we have seen, it is a
consequence of (\ref{e11}).\\ \\
In String Theory this is an expression of the duality relation,
$$R \to \alpha'/R$$
This is symptomatic of the fact that we cannot go down to arbitrarily small
spacetime intervals, below the Planck scale in this case (Cf.ref.\cite{r14}).\\
In the Quantum Mechanical Kerr-Newman Black Hole model of the electron, on the
contrary, we are at Compton scale, and the effect of (\ref{e18}) is precisely
that seen in point 1 above: We go from the electric charge $e$ to the monopole,
as in the Olive-Montonen duality\cite{r35}, (Cf.also ref.\cite{r14}).\\\ \\
{\large {\bf (iv) Spin}}\\ \\
One could argue that the non commutative relations (\ref{e11}) are an expression
of Quantum Mechanical spin. To put it briefly, for a spinning particle the non
commutativity arises when we go from canonical to covariant position variables.
Zakrzewsk\cite{r36} has shown that we have the Poisson bracket relation
$$\{x^j, x^k\} = \frac{1}{m^2}  R^{jk}, (c = 1),$$
where $R^{jk}$ is the spin. The passage to Quantum Theory then leads us back
to the relation (\ref{e11}).\\
Conversely it was shown that the relations (\ref{e11}) imply Quantum Mechanical
spin\cite{r5}. Another way of seeing this is to observe as noted in
(\ref{e13}) that (\ref{e11}) implies
that $y = \alpha \hat p_y ,$ where $\alpha$ is a dimensional constant viz
$[T/M]$ and $\hat p_y$ is the analogue of the momentum, but with the Planck
constant replaced by $l^2$. So the spin is given by
$$| \vec r \times \vec p | \approx 2 x p_y \sim 2 \alpha^{-1} l^2 =
\frac{1}{2} \left(\frac{\hbar}{m^2c^2}\right)^{-1} \times \frac{h^2}{m^2c^2} =
\frac{\hbar}{2}$$
as required.\\ \\
{\large {\bf (v) Extremal Black Holes}}\\ \\
Going back to the relation (\ref{e7}), we can see that if
$$a = \frac{\hbar}{Mc} \sim \frac{GM}{c^2}$$
then we are at the Planck scale and have a Planck mass Schwarzchild Black Hole.
The purely Quantum
Mechanical Compton length equals the classical Schwarzchild radius.\\
Also if,
\begin{equation}
Q \sim Mc^2,\label{e19}
\end{equation}
while at the same time the particlee has no spin, so that $a = 0$, we recover
a Schwarzchild Black Hole. We observe that if the mass $M \sim$ electron mass
then the charge $Q$ from (\ref{e19}) turns out to be $\sim 1000 e$, as in the
case of the monopole.\\
Interestingly these parameters also fit a neutrino, whose mass, as recent
experiments indicate is given by
\begin{equation}
m \leq 10^{-8} m_e\label{e20}
\end{equation}
It was further argued\cite{r37} that a neutrino could in principle have an
electric charge, a millionth that of the electron, while, as the neutrino
has no Compton wavelength we can apply in principle equation (\ref{e19}).
(\ref{e20}) coupled with this and with the above electric charge shows that indeed
equation (\ref{e19}) is satisfied.\\
Such particles however have a very high Bekenstein temperature
$$\sim 10^{-7} \left(\frac{M_0}{M}\right) K,$$
$M_0$ being the solar mass and would disintegrate into gamma rays within about
$10^{-23} M^3 secs$. So these extremal Black Holes would not be detectable, but
this could nevertheless provide a rationale for the puzzling cosmic gamma ray
emissions.\\ \\
{\large {\bf (vi) Spacetime}}\\ \\
We have seen that the spacetime given by (\ref{e11}) is radically different from
its usual description. Infact the usual spacetime is a sort of a stationary spacetime,
a low energy approximation, as will be clear by the following argument. We
start with the Nelsonian theory in which there is a complex velocity potential
$V - \imath U$, due to a double Weiner process. This has been shown to lead
to the usual Quantum Mechanical description\cite{r38}. Indeed the diffusion
equation,
$$\Delta x \cdot \Delta x = \frac{h}{m}\Delta t \equiv \gamma \Delta t$$
can also be written as
$$m \frac{\Delta x}{\Delta t} \cdot \Delta x = h = \Delta p \cdot \Delta x$$
which is the usual Heisenberg description.\\
Using the $WKB$ approximation, the Nelsonian wave function
$$\psi = \sqrt{\rho}e^{(\imath/\hbar)s}$$
becomes
$$(p_x)^{-\frac{1}{2}} e^{\frac{\imath}{\hbar}\int p(x)dx}$$
whence
\begin{equation}
\rho = \frac{1}{p_x}\label{e21}
\end{equation}
In this case the condition $U \approx 0$ gives
$$v \cdot \nabla ln (\sqrt{\rho}) \approx 0$$
that is the probability densityh $\rho$ and hence from (\ref{e21}) the
momentum varies very slowly with $x$.\\
The continuity equation now gives
$$\frac{\partial \rho}{\partial t} + \vec \nabla (\rho \vec v) = \frac{\partial \rho}
{\partial t} = 0$$
which shows that $\rho$ is independent of $t$ also\cite{r18}. This is a scenario
of, strictly speaking, a single particle universe, without environmental effects,
a scenario which is an approximation valid for small incremental changes. (The
more physical scenario takes all the particles in the universe into account,
leading to what may be called stochastic holism\cite{r32}). In this case, we can take limits
to vanishing spacetime intervals, as in the usual theory (Cf. 't Hooft loc. cit).
Spacetime in this description is a differentiable manifold and instead of the
relations (\ref{e11}), spacetime is commutative. Effectively we are neglecting
$l^2$. This has been the backbone of twentieth century physics.\\
On the other hand according to Witten\cite{r39}, "String Theory is a part of
twenty-first century physics that fell by chance into the twentieth century."
It does appear that non commutative fuzzy spacetime is a paradigm for the
twenty-first century.

\end{document}